# Pyroelectric Effect Induced by the Built-in Field in the *p-n* Junction of the Quantum Paraelectric PbTe: Experimental Study

A. V. Butenko, V. Sandomirsky, R. Kahatabi, and Y. Schlesinger

Department of Physics, Bar-Ilan University, Ramat-Gan 52900, Israel

Z. Dashevsky and V. Kasiyan

Materials Engineering Department, Ben-Gurion University of the Negev, P.O.B. 653, Beer-Sheva 84105, Israel

## Abstract

We report here the first observation of a pyroelectric effect in a non-polar semiconductor. This effect originates in the temperature dependent electric dipole of the *p-n* junction. The junction was illuminated by a chopped $CO_2$ laser beam, and periodic and single-pulse pyroelectric signals were observed and measured as a function of temperature, reverse bias voltage and chopper frequency. The measured pyroelectric coefficient is $\simeq 10^{-3}$ μC/cm$^2$K in the region of 40–80 K. The theoretical model describes quantitatively all experimental features. The time evolution of the temperature inside the junction region was reconstructed.

The standard pyroelectric effect (PE) is usually observed only in crystals possessing polar axis symmetry [1]. However, in any non-polar solid, a built-in electric field, such as in the semiconductor barrier structures: *p-n* junctions (PNJ), Schottky contacts, heterojunctions etc. [2], creates a polar direction, or an electric dipole moment. If the dipole moment of the depleted area of a *p-n* junction varies with temperature, it can generate a pyroelectric effect, the junction barrier pyroelectricity, JBP. The magnitude of the JBP effect will be large, provided that the static dielectric constant $\varepsilon$ of the semiconductor depends strongly on temperature. Quantum paraelectrics, such as $SrTiO_3$ [3] or the narrow-gap semiconductor PbTe [4, 5] possess this property.

For the observation of JBP, high-quality diodes have been prepared on a rectangular slice, cut from an ingot of *p*-PbTe single crystal, grown by the Czochralski method. The acceptor concentration was $10^{18}$ $cm^{-3}$. At 80 K the whole diffusion length was ≈ 30 μm, and the mobility was ≈ 14500 $cm^2$/Vs at 80 K. The PbTe absorption coefficient at the wave length of 10.5 μm, with a carrier concentration of $10^{17} - 10^{18}$ $cm^{-3}$, is $\alpha \approx 10^{-2}$ $cm^{-1}$ in the temperature interval of 20 – 200 K [6].

The *n*-region was created by thermodiffusion of In from $In_4Te_3$ gas phase. The junction was formed at a depth of ≈ 70 μm below the exposed surface. The details of the sample preparation methodology, and the detailed account of the characterization of these diodes, will be published elsewhere [7, 8].

The current-voltage characteristics were fitted by the Shockley formula [2]. The fitting resulted in the ideality factor n of ≈ 1.5 ÷ 2, and a saturation current density of ≈ $10^{-5}$ A/$cm^2$ at 80 K. The junctions were found to be linearly graded [2]. The temperature dependence of the dielectric constant $\varepsilon(T)$, derived from the temperature dependence of the junction capacitance, is fitted by Barrett's formula [9]

$$\epsilon = \frac{1.36 \cdot 10^5}{36.14 \cdot \coth(36.14/T) + 49.15}. \tag{1}$$

The pyroelectric signal (PES) was measured using the experimental setup shown schematically in Fig. 1. The PES temporal variation has been measured by a Tektronix Differential Preamplifier ADA400A with display on Tektronix e-scope TDS 3054B. The PES under bias, $V_b$, has been measured by an EG&G Princeton Applied Research Lock-in Amplifier 5209, in parallel with a Keithley Sourcemeter 2410 for applying $V_b$. The output signal was displayed on the e-scope. The signal was found to depend linearly on the laser beam intensity $P$. The chopper frequency $f$ was varied from 4 Hz to 2000 Hz. The width of the chopper slit was much larger than the lateral width of the junction. Therefore, the time dependence of the light intensity on the illuminated area was trapezoid shaped, with short rise- and fall-times ($t_1$), and a long period of constant illumination/darkness ($t_2$).

The excitation was applied in two modes: (1) A periodic mode (PPES) with equal durations of illumination and darkness; the period was shorter then the characteristic times of the temperature relaxation ($t^*$) and $\tau_e = RC$ of the diode (see below). (2) A single pulse mode (SPES) with a long darkness period, so that the temperature and $RC$ relaxations have been completed before the occurrence of the next pulse.

The measurements were performed over the temperature interval 12 − 130 K. The samples were placed in a closed cycle He-gas refrigerator cryostat, in a vacuum of about $10^{-7}$ torr. The temperature was controlled and stabilized by a LakeShore DRC-91CA Temperature Controller.

The JBP theory is constructed, akin to the standard theory of PE for thin pyroelectric films [10-12], with specific attention to the junction depleted region [2]. The PE theory is based on the simultaneous solution of a pair of equations. One is the thermal balance equation, and the other is the Kirchhoff equation for the quasi-stationary currents. The thermal balance equation reads

$$C_T\frac{dT}{dt} + G_T \bullet \Delta T = A(t) \bullet P, \tag{2}$$

where $C_T$ is the total heat capacity of the junction, $G_T$ is the heat transfer coefficient, $A(T)$ is the illuminated sample area. Here $T$ is the junction temperature, and $T_0$ is the temperature as set by the temperature controller (Fig.1). The term $G_T{\cdot}\Delta T$ describes the Newton heat transfer.

The equation for the current reads

$$\frac{d[C(V_t) \bullet V_t]}{dt} + \frac{U}{R(V_t)} = 0, \tag{3}$$

where $V_t = V_0 - V_b + U$, $V_0$ is the built-in barrier potential, $U$ is the PES. Typical value of $V_0$ is a few tens of mV, and of $V_b$ a few hundreds of mV. Since $U$ (of the order of μV) is much less then both $V_0$ and $|V_b|$, one can expand the terms in Eq. (3) into a power series in $U$, keeping linear terms only. Then, designating $V = V - V_b$, we obtain

$$\frac{dU}{dt} + \frac{U}{R(V) \bullet C(V)} = \Lambda \bullet \frac{dT}{dt}; \Lambda = -V \bullet \frac{d\ln C(V)}{dT}; \lambda = \frac{\Lambda}{A_e} \tag{4}$$

where $A_e$ is the area of the junction. Thus, $\Lambda$ is the PE coefficient of JBP, and $\lambda$ is the usual PE coefficient. Eq. (4) shows that JBP is zero when the PNJ barrier vanishes ($V_0 = 0$). From Eq. (4) also follows that the value of $\Lambda$ depends on doping (via $V_0$), on the doping gradient (via $C$), on $\varepsilon(T)$ (via $C$), on $V_b$ and on $T_0$. The graph of $\lambda$ vs. $T$ is shown in Fig. 2; $\lambda$ is negative in accord with $d\varepsilon/dT < 0$ (Eq. (1)).

The solution of the pair of equations (2) and (4) is determined by three parameters: $t^* = C_T / G_T$, $p = t^*/RC$ and $\Lambda$.

The left-hand side of Eq. (4) consists of two currents

$$I_1 = C(V)\frac{dU}{dt} \quad \text{and} \quad I_2 = \frac{U}{R(V)}. \tag{5}$$

$I_1$ is the displacement current due to the change of the PNJ dipole moment; $I_2$ is the conduction current, screening the non-equilibrium PNJ polarization.

Equation (4) can be applied also independently on Eq. (2), to deduce $\Delta T(t)$ in the PNJ for an arbitrary, JBP generating, source. Indeed, if the PES $U(t)$ is measured, than Eq. (4) gives d$T$/d$t$

$$\frac{dT}{dt} = \frac{C(V)}{\Lambda}\frac{dU}{dt} + \frac{U}{R(V)\cdot\Lambda} \tag{6}$$

Integrating Eq. (6) one can find $\Delta T(t)$.

Figure 3 shows an example of the measured and calculated time variation of PPES at $T = 80.1$ K, at different frequencies and at $V_b = 0$. The calculated shape of PPES agrees well with the experiment.

The main features of these time dependent variations are:

(1) The magnitude of PPES is $\sim 10 - 20$ μV, and PPES exhibits the typical "pyroelectric" shape.

(2) PPES shape varies with frequency rather markedly.

(3) PPES changes its sign in the dark phase.

(4) The PPES-amplitude drops with increasing temperature, since the built-in barrier $V_0$ vanishes at higher temperatures and $\Lambda \to 0$.

(5) The PPES-amplitude increases with the reverse bias, as $\Lambda$ increases with $V_b$ (Fig.2).

The single pulse SPES behaves similarly.

Example of the derived $\Delta T(t)$ inside PNJ, and of the temporal variation of the currents $I_1$ and $I_2$ is illustrated in Fig. 4. These plots show that: (1) The amplitude of $\Delta T(t)$ is of the order of $\approx 10$ mK at low temperatures and reaches $\approx 1$ K at higher temperatures. The latter results in a decrease of the thermodiffusivity at high temperature. Respectively, the heat transfer between the PNJ and the adjoining medium decreases. (2) For the same reason, $\mathrm{d}T/\mathrm{d}t$ increases also with temperature. (3) The current amplitude is $10^2 - 10^3$ nA, and increases with temperature. $I_2 \ll I_1$ at low temperature, and $I_2 \gg I_1$ at higher temperature. That results in the exponential decrease of the PNJ resistance. (4) The screening current lags after the displacement current, expressing the fact that it is the change of polarization $I_1$ that initiates $I_2$.

In conclusion, a pyroelectric effect in a PNJ of a non-polar semiconductor was observed for the first time. Contrary to a standard PE, the JBP in PbTe increases with decreasing temperature. The theoretical formulation describing the JBP has been successfully applied. The PES theory allows deducing the time variation of the temperature in the PNJ. Other quantum paraelectrics, e.g. $SrTiO_3$, having a significantly larger dielectric constant, and stronger temperature dependence than that of PbTe, are promising candidates for a large JBP and are presently investigated. We observed also a similar magnitude JBP in a Schottky contact barrier of In–*p*-PbTe.

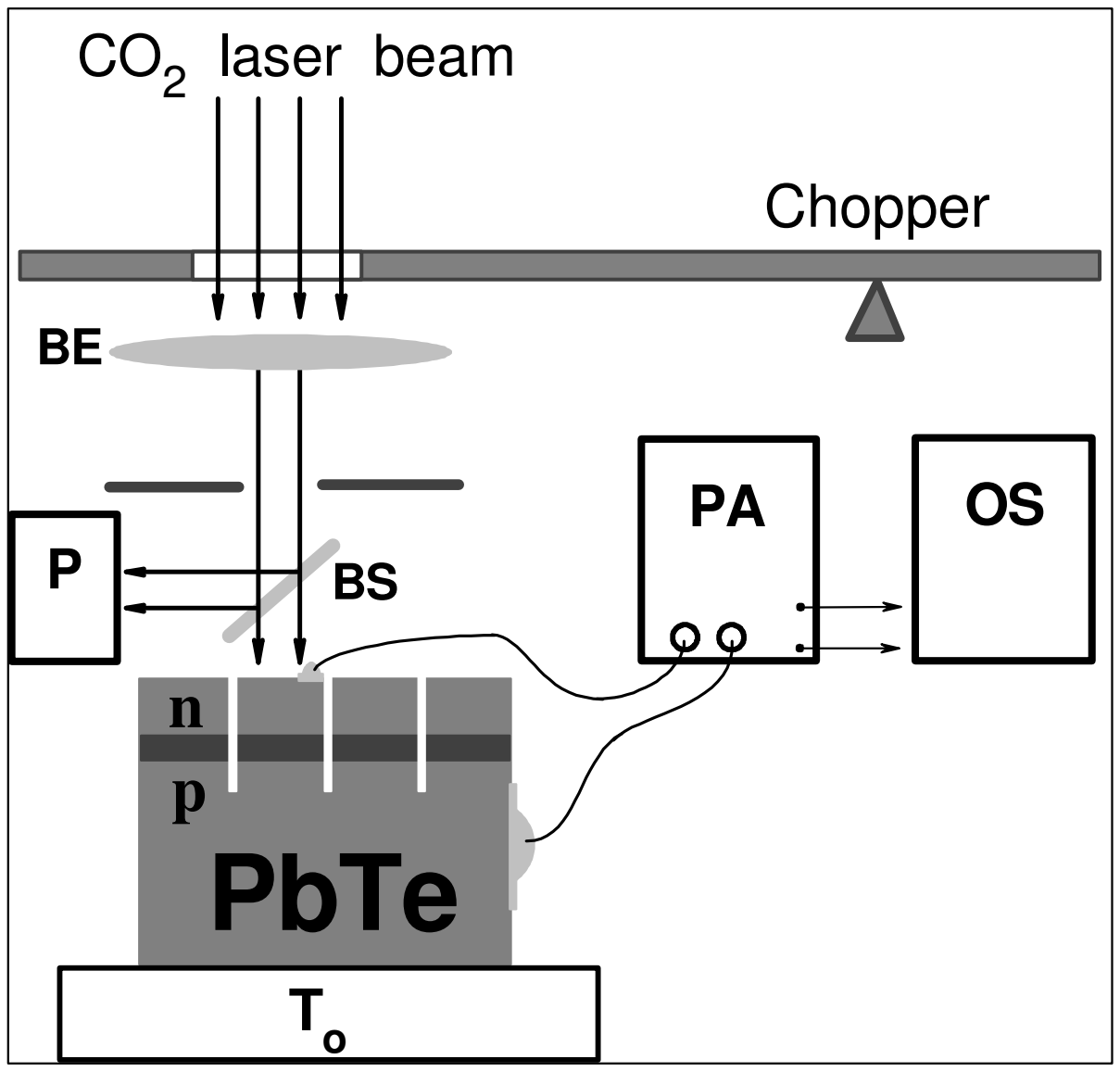

Fig. **1** Schematic drawing of the experimental set-up.

BE – beam expander; BS - beam splitter; P – light intensity monitor; PA – preamplifier; OS –oscilloscope; $T_0$ – sample temperature set by the temperature controller.

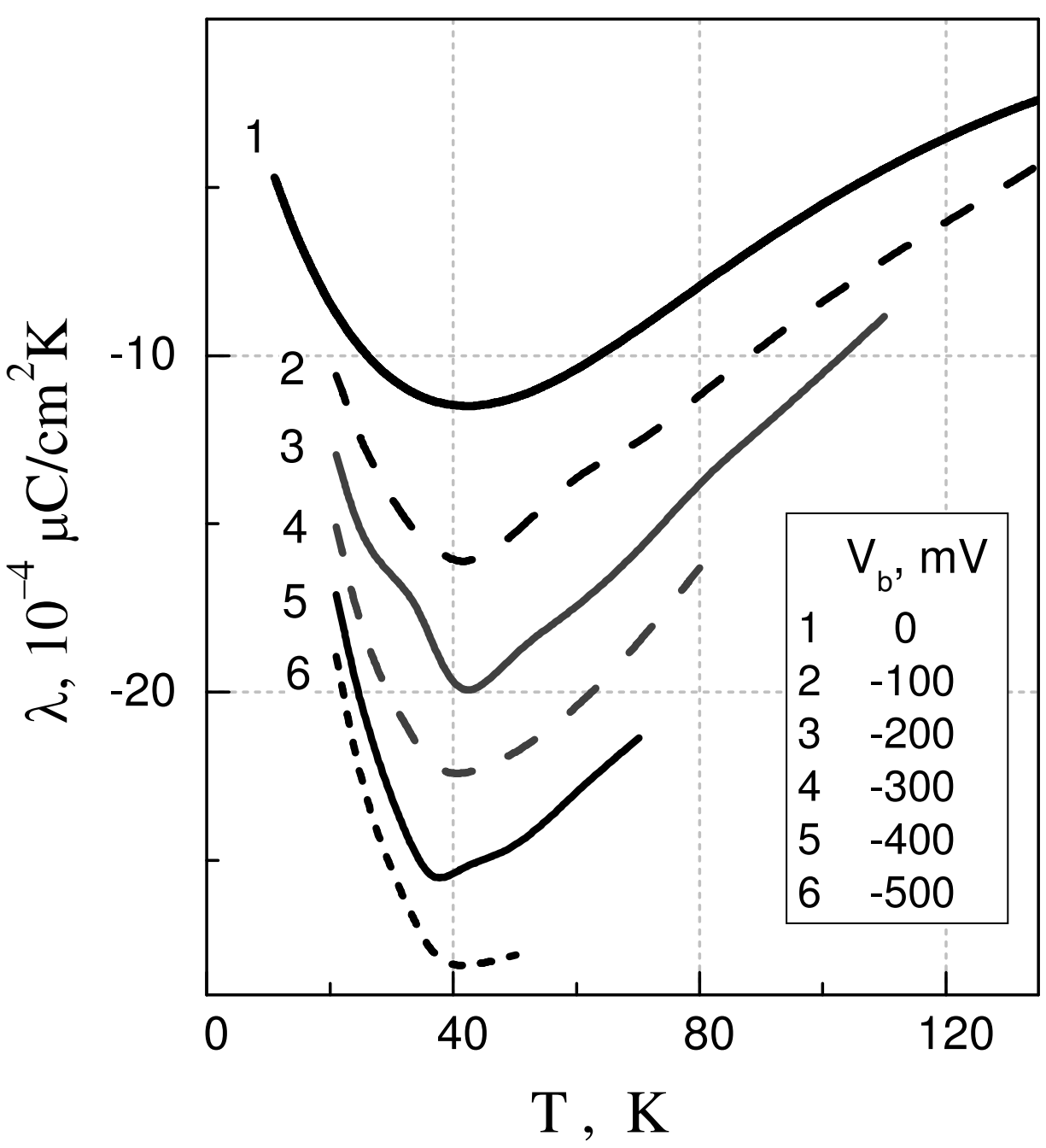

Fig. **2** The temperature dependence of the PE coefficient at different bias voltages.

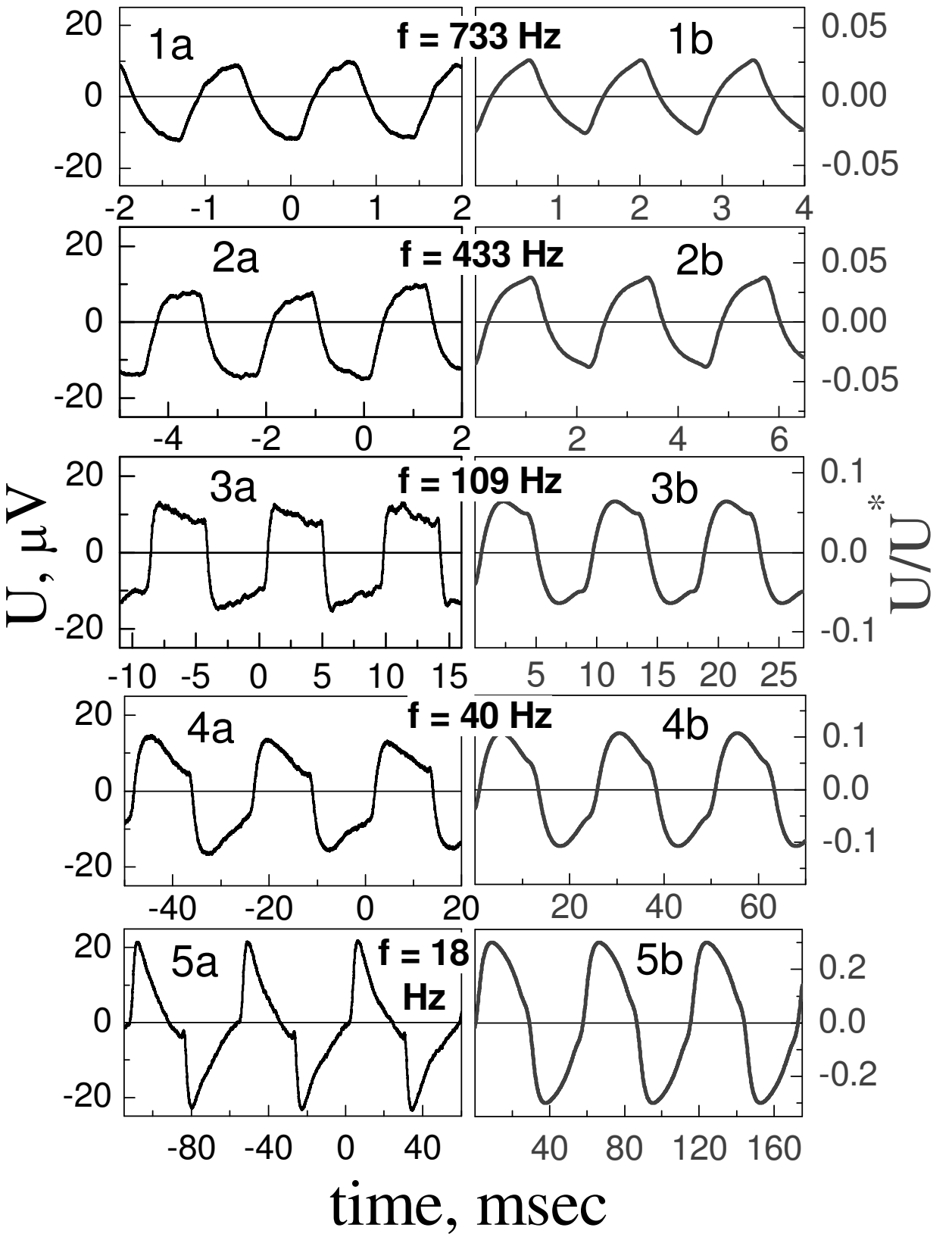

Fig. **3** The pyroelectric signal at 80.1 K, and different chopper frequencies.

a – experiment, b – theoretical calculation (in normalized units).

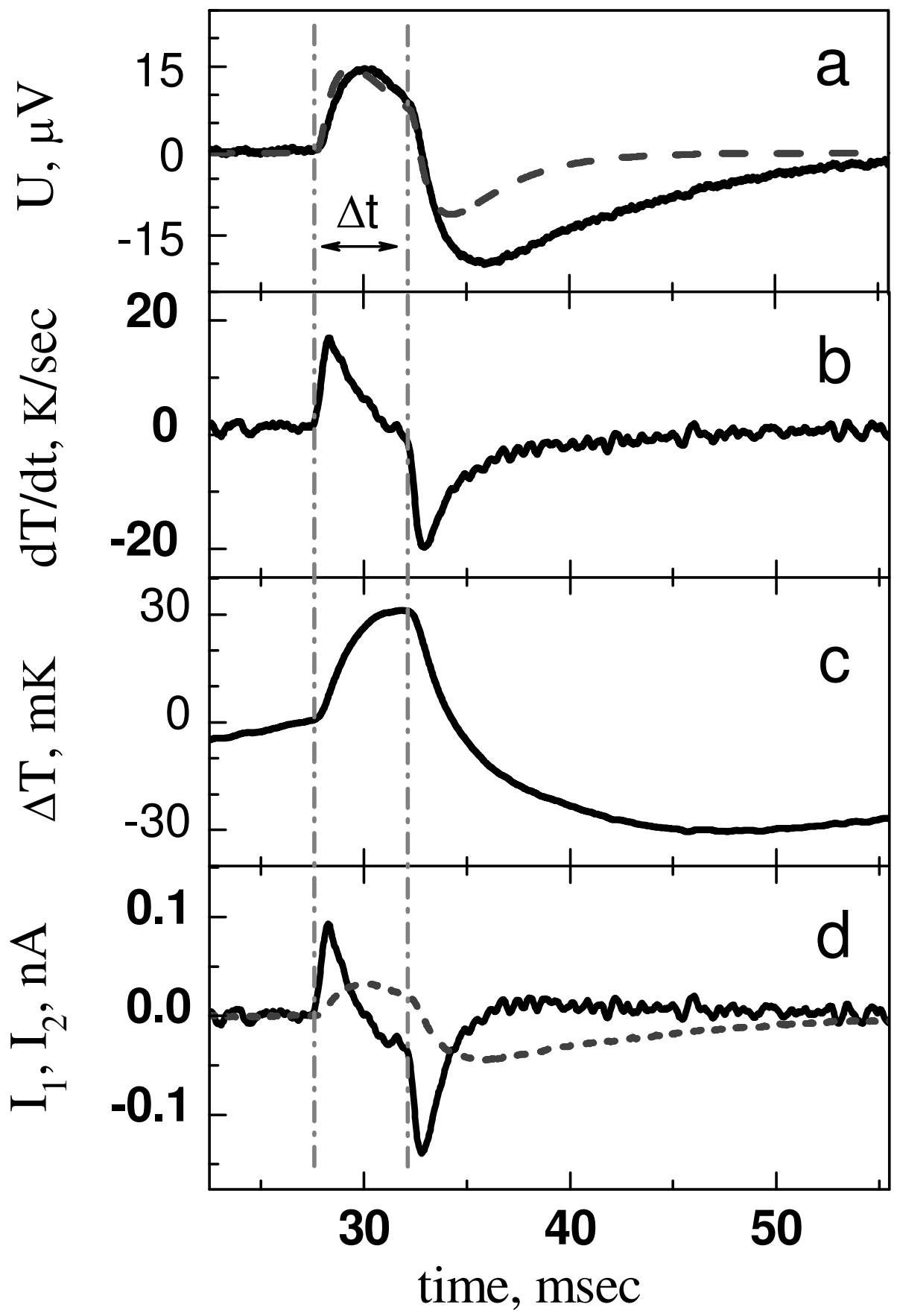

Fig. **4** The single-pulse pyroelectric response at 60.2 K.

(a) – The pyroelectric signal (solid-line:experimental, dashed-line: calculated); (b) – The rate of change of the temperature in the junction region; (c) – The deduced time evolution of the temperature in the junction region; (d) - The displacement ($I_1$, solid-line) and the conduction ($I_2$, dashed-line) currents.

$\Delta t$ = 4.6 msec, the time between pulses is 50 msec.